\documentstyle[11pt] {article}
\begin{document}

   \title{Relativistic quantum field inertia and vacuum field noise spectra}

   \author{Haret C. Rosu \and Inst. de F\'{\i}sica, Univ. de
 Guanajuato, A. P. E-143, Le\'on, Gto, Mexico}


   \maketitle

\hskip3cm Int. J. Theor. Phys. {\bf 39} (2000)

   \begin{abstract}

The main theme of this survey is the equivalence statements for quantum
scalar field vacuum states that have been recognized over the last couple
of decades as a powerful line of reasoning when discussing the highly
academic thermal-like Hawking effect and Unruh effect.
An important ingredient in this framework is the concept of vacuum field
noise spectrum by which one can obtain
information about the curvature
invariants of classical worldlines (relativistic classical trajectories).
It is argued, in the spirit of the free fall type universality, that the
preferred quantum field vacua with respect to accelerated worldlines should
be chosen in the class of all those possessing stationary spectra for their
quantum fluctuations. For scalar quantum field vacua there are six stationary
cases as shown by Letaw some time ago, these are reviewed here. However, the
non-stationary vacuum noises are not out of reach and can be processed by a
few mathematical methods that are mentioned as well. Since the information
about the kinematical curvature invariants of the worldlines is of radiometric
origin, hints are given on a more useful application of such an academic
formalism to radiation and beam radiometric standards at high energy
accelerators and in astrophysics. The survey ends up with a quick look to
related axiomatic quantum field topics and a few other recent works.

               
   \end{abstract}

%


\section{Introduction}

The legendary {\em gedanken} discovery of {\em classical free
fall universality}
by Galilei \cite{gaga} in the first instants of modern science is now, for
everybody, an early textbook exciting story
(first actual experiments in June 1710 at St. Paul's in London
by Newton). Starting
with the neutron beam experiments of
Dabbs {\em et al} \cite{dabbs} in 1965
non-relativistic {\em quantum free falls} have also been of much
interest. As known, `free falls' of quantum wavefunctions (wavepackets),
i.e. Schroedinger solutions in a homogeneous gravitational field,
are mass dependent and therefore closer to Aristotle's fall. Thus,
a reset of the quest for the universality features of free
fall type phenomena in the quantum realm has emerged in recent epochs.
Moreover, at the present time, there are interesting insights in the
problem of {\em relativistic quantum field inertia},
which have been gained as a consequence of the
Hawking effect \cite{heff}
and the Unruh effect \cite{ueff}. This substantially helped to display
the `imprints' of
gravitation in the relativistic quantum physics \cite{ug}.
Natural questions in this context on which I hope to be sufficiently
informal during this work could be
(i) What does really mean `free fall'
in relativistic quantum field theories ?
(ii)
How should one formulate EPs for quantum field states ?
(iii) What are the restrictions on quantum field states imposed by the EPs ?

\noindent
The method of quantum detectors proved to be very useful for the
understanding of
the quantum field inertial features. New ways of thinking of
quantum fluctuations have been promoted and new
pictures of the vacuum states have been provided, of which the landmark one is
the heat bath interpretation of the Minkowski vacuum state from
the point of view of a uniformly accelerating non-inertial quantum detector.
Essentially, simple, not to say toy, model particles
(just two energy levels separated by $E$ and monopole form factor) commonly
known as Unruh-DeWitt (UDW) quantum detectors
of uniform, one-dimensional proper acceleration $a$ in Minkowski vacuum are
immersed in a scalar quantum field `heat' bath of temperature
\begin{equation} \label{1}
T_{a}=\frac{\hbar}{2\pi c k}\cdot a~,
\end{equation}
where $\hbar$ is Planck's barred constant, $c$ is the speed of light in
vacuum, and $k$ is Boltzmann's constant. A formula of this type has been
first obtained by Hawking in a London Nature Letter of March 1974
on black hole explosions \cite{heff}, then in 1975 by
Davies in a moving mirror model \cite{deff}, and finally settled by Unruh
in 1976 \cite{ueff}.
For first order corrections to this formula one can see works
by Reznik \cite{rez}.
This Unruh temperature is
proportional to the lineal uniform acceleration, and the scale of such
noninertial quantum field `heat' effects with respect to the acceleration one
is
fixed by the numerical values of universal constants to the very low value of
$4\times 10^{-23}$ in cgs units). In other words, the huge acceleration of
$2.5 \times 10^{22}$ ${\rm cm/s}^2$ can produce a black body spectrum of
only 1 K.
In the (radial) case of Schwarzschild black holes, using the surface gravity
$\kappa=c^{4}/4GM$ instead of $a$, one immediately gets the formula for their
Hawking temperature, $T_{\kappa}$.
In a more physical picture, the Unruh quantum field heat reservoir is
filled with the so-called Rindler photons
(Rindler quasi-particles), and
therefore the quantum transitions are to be described as absorptions or
emissions of the Rindler reservoir `photons'.

\noindent
I also recall that according to an idea popularized by Smolin \cite{smo},
one can think of
{\em zero-point fluctuations}, {\em gravitation} and {\em inertia} as the
only three {\em universal} phenomena of nature. However, one may also think
of inertia as related to those
peculiar collective, quantum degrees of freedom which are the vacuum
expectation values (vev's) of Higgs fields. As we know,
these vev's do not follow from the fundamentals of quantum theory.
On the other hand, one can find
papers claiming that inertia can be assigned to a Lorentz type force
generated by electromagnetic zero-point fields \cite{hrp}.
Moreover, it is quite well known the
{\em Rindler condensate} concept of Gerlach \cite{ger}. Amazingly,
one can claim that there exist completely coherent zero-point condensates,
like the Rindler-Gerlach one, which entirely mimick the Planck spectrum,
without any renormalization, as the case is for the Casimir effect.

\noindent
In this work, I will stick to the standpoint based on the
well-known concept
of vacuum field noise (VFN) \cite{tak}, - or vacuum excitation
spectrum from the point of view of quantum UDW detectors - because in my
opinion this not only provides a
clear origin of the relativistic thermal effects, it avoids at
the same time uncertain generalizations, and also helps one of my purposes
herein. This is to shed more light on the connection between the stationary
VFNs and the equivalence principle statements for scalar field theories.

\section{Survey of quantum detector EPs}

\noindent
The Unruh picture can be used for interpreting
Hawking radiation in Minkowski space \cite{jac}. In order to do that,
one has to consider the generalization(s) of the EP to
quantum field processes. A number of authors have discussed this important
issue with various degree of detail and meaning and with some debate \cite{1}.
Nikishov and Ritus \cite{nr} raised the following objection to the
heat bath concept. Since absorption and emission processes occur in finite
space time regions, the application of the
local principle of equivalence requires a constant acceleration
over those regions. However, the space-time extension of the quantum
processes are in general of the order of inverse acceleration. In Minkowski
space it is not possible to create homogeneous and uniform
gravitational fields having accelerations of the order of $a$ in
spacetime domains of the order of the inverse of $a$.   

\noindent
Grishchuk, Zel'dovich, and Rozhanskii, and also
Ginzburg and Frolov wrote extensive reviews on the formulations
of QFEP \cite{1}. One should focus on the response functions of
quantum detectors, in particular the UDW two-level monopole
detector in stationary motion.
In the asymptotic limit this response function is the integral of the
quantum noise power spectrum. Or, since the derivative of the response
function is the quantum transition rate, the latter is just the measure of
the vacuum power spectrum along the chosen trajectory (worldline) and in
the chosen initial (vacuum) state. This is valid only in the asymptotic limit
and more realistic cases require calculations in finite time
intervals \cite{ss}. Denoting
by $R_{M,I}$, $R_{R,A}$, and $R_{M,A}$ the detection
rates with the first subscript corresponding to the vacuum (either Minkowski
or Rindler) and the second subscript corresponding to either inertial or
accelerating worldline, one can find for the UDW detector in a
scalar vacuum that $R_{M,I}=R_{R,A}$ expressing the dissipationless
character of the vacuum fluctuations in this case, and a thermal factor
for $R_{M,A}$ leading to the Unruh heat bath concept. In the case of
a uniform gravitational field, the candidates for the vacuum state are
the Hartle-Hawking ($HH$) and the Boulware ($B$) vacua. The $HH$ vacuum is
defined by choosing incoming modes to be those of positive frequency with
respect
to the null coordinate on the future horizon and outgoing modes as
positive frequency ones with respect to the null coordinate on the past
horizon, whereas the $B$ vacuum has the positive frequency modes with
respect to the Killing vector which makes the exterior region static.
For an ideal, uniform gravitational field the $HH$ vacuum can be thought of
as the counterpart of the Minkowski vacuum, while the $B$ vacuum is the
equivalent of the Rindler vacuum. Then, the QFEP can be formulated in one of
the following ways

\vskip 0.3cm

\noindent
{\bf Quantum detector-QFEP: $HH-M$ equivalence}

\noindent
{\sl i) The detection rate of a free-falling UDW detector in the
HH vacuum is the same as that of an inertial UDW detector in the M vacuum.

\noindent
ii) A UDW detector at rest in the HH vacuum has the same DR as a uniformly
accelerated detector in the M vacuum.}

\vskip 0.3cm

\noindent
{\bf Quantum detector-QFEP: $B-R$ equivalence}

\noindent
{\sl iii) A UDW detector at rest in the B vacuum has the same detection rate
as a uniformly accelerated detector in the R vacuum.

\noindent
iv) A free-falling UDW detector in the B vacuum has the same detection rate
as an inertial detector in the R vacuum.}

\vskip 0.3cm

\noindent
Let us record one more formulation due to Kolbenstvedt \cite{1}

\vskip 0.3cm

\noindent
{\bf Quantum detector-QFEP: Kolbenstvedt}

\noindent
{\sl A detector in a gravitational field and an accelerated detector will
behave in the same manner if they feel equal forces and perceive radiation
baths of identical temperature}.

\vskip 0.1cm

\noindent
In principle, since the Planck spectrum is Lorentz invariant
(and even conformal invariant) its presence in equivalence statements
is easy to accept if one reminds that Einstein EP requires local Lorentz
invariance. The linear connection between `thermodynamic' temperature and
one-dimensional, uniform, proper acceleration, which is also
valid in some important gravitational contexts (Schwarzschild black holes,
de Sitter cosmology), is indeed a
fundamental relationship, because it allows for an absolute meaning of
quantum field effects in such {\em ideal} noninertial frames, as soon as
one recognize thermodynamic temperature as
the only {\it absolute}, i.e., fully {\it universal} energy type physical
concept.

\section{The six types of stationary scalar VFNs}  

\noindent
In general the scalar quantum field vacua
are not stationary stochastic processes (abbreviated as SVES)
for all types of classical trajectories on which the UDW detector moves.
Nevertheless, the lineal
acceleration is {\em not} the only case with that property as was shown by
Letaw \cite{let} who
extended Unruh's considerations, obtaining six types
of worldlines with SVES for UDW detectors (SVES-1 to SVES-6, see below). These
worldlines are solutions of some generalized Frenet equations
on which the condition of constant curvature invariants is imposed,
i.e., constant curvature $\kappa$, torsion $\tau$, and
hypertorsion $\nu$, respectively. Notice that one can employ other frames
such as the Newman-Penrose spinor formalism as recently
did Unruh \cite{nepe} but the Serret-Frenet one is in overwhelming use
throughout physics. The six stationary cases are the following

\bigskip
\underline{1. $\kappa =\tau=\nu=0$},
(inertial, uncurved worldlines). SVES-1 is a trivial cubic spectrum
\begin{equation}
S_1(E)=\frac{E^3}{4\pi ^2}
\end{equation}
i.e., as given by a vacuum of zero point energy per mode $E/2$
and density of states $E^2/2\pi ^2$.

\bigskip
\underline{2. $\kappa \neq 0$, $\tau=\nu=0$},
(hyperbolic worldlines). SVES-2 is Planckian allowing the
interpretation of $\kappa/2\pi$ as `thermodynamic' temperature. In the
dimensionless variable $\epsilon _{\kappa}=E/\kappa$ the vacuum spectrum reads
\begin{equation}
S_2(\epsilon _{\kappa})
=\frac{\epsilon _{\kappa}^{3}}{2\pi ^2(e^{2\pi\epsilon _{\kappa}}-1)}
\end{equation}

\bigskip
\underline{3. $|\kappa|<|\tau|$, $\nu=0$, $\rho ^2=\tau ^2-\kappa ^2$},
(helical worldlines). SVES-3 is an analytic function
corresponding to case 4 below only in the limit $\kappa\gg \rho$
\begin{equation}
S_3(\epsilon _{\rho})\stackrel{\kappa/\rho\rightarrow \infty}
{\longrightarrow} S_4(\epsilon _{\kappa})
\end{equation}
Letaw plotted the numerical integral $S_3(\epsilon _{\rho})$,
where $\epsilon _{\rho}=E/\rho$ for various values of $\kappa/\rho$.

\bigskip
\underline{4. $\kappa=\tau$, $\nu=0$},
(the spatially projected worldlines are the semicubical parabolas
$y=\frac{\sqrt{2}}{3}\kappa x^{3/2}$ containing a
cusp where the direction of motion is reversed). SVES-4 is analytic, and
since there are two equal curvature invariants one can use the
dimensionless energy variable $\epsilon _{\kappa}$. 
\begin{equation}
S_{4}(\epsilon _{\kappa})= \frac{\epsilon _{\kappa}^{2}}{8\pi ^2 \sqrt{3}}
e^{-2\sqrt{3}\epsilon _{\kappa}}
\end{equation}
It is worth noting that $S_4$ is rather close to the Wien-type spectrum
$S_{W}\propto\epsilon ^3e^{- {\rm const.}\epsilon}$.

\bigskip
\underline{5.  $|\kappa|>|\tau|$, $\nu=0$, $\sigma ^2=\kappa ^2-\tau ^2$},
(the spatially projected worldlines are catenaries, i.e., curves of the type
$x=\kappa \cosh (y/\tau)$). In general, SVES-5 cannot be found
analitically. It is an intermediate case, which
for $\tau/\sigma\rightarrow 0$ tends to SVES-2,
whereas for $\tau/\sigma\rightarrow\infty$ tends toward SVES-4
\begin{equation}
S_2(\epsilon _{\kappa})
\stackrel{0\leftarrow \tau/\sigma}{\longleftarrow}
S_5(\epsilon _{\sigma})\stackrel{\tau/\sigma\rightarrow \infty}
{\longrightarrow}S_4(\epsilon _{\kappa})
\end{equation}

\bigskip
\underline{6. $\nu\neq 0$},
(rotating worldlines uniformly accelerated normal to their plane of rotation).
SVES-6 forms a two-parameter set of curves. These trajectories are
a superposition of the constant linearly accelerated motion and uniform
circular motion. The corresponding vacuum spectra have not been calculated
by Letaw even numerically. 

Thus, only the hyperbolic worldlines having just one nonzero curvature
invariant allow for a Planckian SVES and for a strictly one-to-one
mapping between the curvature invariant $\kappa$ and the `thermodynamic'
temperature.
On the other hand, in the stationary cases it is possible to determine at
least approximately the curvature invariants, that is the
classical worldline on which a quantum particle moves, from measurements
of the vacuum noise spectrum.

\section{Preferred vacua and/or high energy radiometric standards}

\noindent
There is much interest in considering the
magnetobremsstrahlung (i.e., not only synchrotron) radiation patterns at
accelerators in the aforementioned perspective \cite{mont} at least since
the works of Bell and collaborators \cite{bell}.
It is in this sense that a sufficiently general and acceptable statement
on the {\em universal} nature of
the kinematical parameters occurring in a few important quantum field model
problems can be formulated as follows

\vskip 0.1cm

\noindent
{\em  There exist accelerating
classical trajectories (worldlines) on which moving ideal (two-level)
quantum systems can detect the scalar vacuum environment as a stationary
quantum field vacuum noise with a spectrum directly related to
the curvature invariants of the worldline, thus allowing for a
radiometric meaning of those invariants}.

\bigskip

\noindent
Although this may look an extremely ideal (unrealistic) formulation
for accelerator radiometry, where the spectral photon flux formula of
Schwinger \cite{sflux} is very effective,
I recall that Hacyan and Sarmiento \cite{hs} developed a formalism similar
to the scalar case to calculate the vacuum stress-energy tensor of the
electromagnetic field in an arbitrarily moving frame and applied it to a
system in uniform rotation, providing formulas for the energy density,
Poynting flux and stress of zero-point oscillations in such a frame.
Moreover, Mane \cite{mane} has suggested the Poynting flux of Hacyan and
Sarmiento to be in fact synchrotron radiation when it is coupled to an
electron.

\noindent
Another important byproduct and actually one of the proposals I put forth
in this essay is the possibility to choose a class of
preferred vacua of the quantum world \cite{pr} as {\em all} those
having stationary vacuum noises with
respect to the classical (geometric) worldlines of {\em constant} curvature
invariants because in this case one may find some
necessary attributes of universality in the more general
quantum field radiometric sense \cite{rad} in which the Planckian Unruh
thermal spectrum is included as a particularly important case.
Of course, much work remains to be done for a more ``experimental"
picture of highly academic calculations in quantum field theory,
but a careful look to the literature
shows that there are already definite steps in this direction \cite{steps}.
One should notice that all the
aforementioned scalar quantum field vacua look extremely ideal from the
experimental standpoint. Indeed, it is known that only strong external fields
can make the quantum electrodynamical vacuum to react and show its physical
properties, becoming similar to a magnetized and polarized medium,
and only by such means one can learn about the physical structure of the
QED vacuum.
Important results on the relationship between Schwinger mechanism
and Unruh effect have been reported in recent works \cite{gab}.

\section{Nonstationary VFNs}

\noindent
Though the nonstationary VFNs do not enter statements of equivalence type
they are equally important.
Since such noises have a time-dependent spectral content one needs joint
time and frequency information, i.e. generalizations of the power spectrum
analysis such as tomographical processing \cite{manko} and wavelet
transform analysis \cite{wavl}.
Alternatively,
since in the quantum detector method the vacuum autocorrelation functions
are the essential physical quantities, and since according to
fluctuation-dissipation theorem(s) (FDT) they are related to the linear
(equilibrium) response functions to an initial condition/vacuum,
more FDT type work, especially its generalization to the out of
equilibrium case \cite{out} will be useful in this framework. One can hope
that effective temperature concepts can be introduced following the
reasoning already developed for systems with slow dynamics
(glasses) \cite{glass}.
In fact, there is some progress due to
Hu and Matacz \cite{hm} in making more definite
use of FDT for vacuum fluctuations.
Very recently, Gour and Sriramkumar \cite{gs} questioned if small particles
exhibit Brownian motion in the quantum vacuum and concluded that even though
the answer is in principle positive the effect is extremely small and thus
very difficult to detect experimentally.

\section{Axiomatic QFEPs}   

\noindent
At the rigorous, axiomatic level, Hessling \cite{hes} published further
results on the algebraic quantum field equivalence principle (AQFEP) due to
Haag and collaborators.
Hessling's formulation is too technical to be reproduced here.
The difficulties are related to the rigorous formulation
of {\em local position invariance}, a requisite of equivalence, for the
singular short-distance behavior of quantum fields, and to the generalization
to interacting field theories.
Various general statements of locality \cite{haa} for linear quantum fields
are important steps toward proper formulations of AQFEP. These are nice but
technical results coming out mainly from
clear mathematical exposition involving algebraic-thermal states, namely
the Kubo-Martin-Schwinger states of Hadamard type.
Hessling's AQFEP formulation is based on the notion of quantum states
{\em constant up
to first order} at an arbitrary spacetime point, and means that for these
states a certain
scaling limit should exist, and moreover a null-derivative condition with
respect to a local inertial system around that arbitrary point is to be
fulfilled for all n-point functions.
For example, the vacuum state of the
Klein-Gordon field in Minkowski space with a suitable {\em scaling function}
fulfills Hessling's AQFEP.
Using as a toy model the asymptotically free $\phi ^3$ theory
in six-dimensional
Minkowski space, Hessling showed that the derivative condition of his AQFEP
is not satisfied
by this interacting quantum field theory, which perturbatively is similar to
quantum chromodynamics. This failing is due to the running coupling constant
that does not go smoothly to zero in the short-distance limit.
If one takes AQFEP or generalizations thereof as
a {\em sine qua non criterium} for physically acceptable quantum field
vacuum states then one has at hand a useful selection guide
for even more complex vacua such as the Yang-Mills one \cite{shif} or those
of quantum gravity \cite{gsqg}.


\noindent
Since the {\em time-thermodynamics} relation
in general covariant theories and the connection with Unruh's temperature and
Hawking radiation are an active area of research due to the remarkable
correspondence between causality and the modular Tomita-Takesaki
theory \cite{crov} it would be interesting to formulate in this context some
sort of AQFEP statement beyond that of Hessling.

\noindent
Finally, the work of Faraggi and Matone \cite{fm} is to be noted, where
a sort of mathematical
equivalence postulate is introduced stating that all physical systems can
be connected by a coordinate transformation to the free system with
vanishing energy, uniquely leading to the quantum analogue of the
Hamilton-Jacobi equation, which is a third-order non-linear differential
equation. The interesting feature of their approach, which they carry on in
both nonrelativistic and relativistic domains, is the derivation of
a trajectory representation of quantum mechanics
depending on the Planck length.

 
\section{Conclusions}                       

\noindent
The first conclusion of this work is that considerations of equivalence type
in quantum field theories may well guide the abstract research in this area
towards the highly required feature of {\em universality},
which being an important
form of {\em unification} is among the ultimate purposes of meaningful
theoretical research. This may go till the act of measuring generic field
operators as was argued by D'Ariano \cite{ume} for the homodyne tomography
technique in quantum optics.

\noindent
The second conclusion refers to the hope that
Hawking and Unruh effects are not only mathematical idealizations.
Especially their vacuum excitation spectrum interpretation
can be used for what one may call high energy kinematical radiometry,
at least as guiding principles in establishing rigorous high energy and
astrophysical radiometric standards. Whether or not Unruh's and Hawking's
effects may really occur \cite{notU} they can be employed as a sort of
standards in relativistic quantum field radiometry.


\end{document}